\documentclass[fleqn,usenatbib]{mnras}

\usepackage{newtxtext,newtxmath}

\usepackage[T1]{fontenc}

\DeclareRobustCommand{\VAN}[3]{#2}
\let\VANthebibliography\thebibliography
\def\thebibliography{\DeclareRobustCommand{\VAN}[3]{##3}\VANthebibliography}

\usepackage{multicol}
\usepackage{graphicx}
\usepackage{subcaption}
\usepackage{amsmath}	

\title[PKS 0605-085 as the origin of the KM3-230213A event]{The blazar PKS 0605-085 as the origin of the KM3-230213A \\ ultra high energy neutrino event}
\author[T.A. Dzhatdoev]
{Timur A. Dzhatdoev$^{1,2}$\thanks{E-mail: timur1606@gmail.com}\\
$^{1}$ Institute for Nuclear Research of the Russian Academy of Sciences, 60th October Anniversary Prospect 7a, Moscow 117312, Russia \\
$^{2}$ Federal State Budget Educational Institution of Higher Education, M.V. Lomonosov Moscow State University, \\ Skobeltsyn Institute of Nuclear Physics (SINP MSU), 1(2), Leninskie gory, GSP-1, 119991 Moscow, Russia}
\date{Accepted XXX. Received YYY; in original form ZZZ}
\pubyear{2025}

\begin{document}
\label{firstpage}
\pagerange{\pageref{firstpage}--\pageref{lastpage}}
\maketitle

\begin{abstract}
The KM3Net Collaboration has recently reported on the observation of a remarkable event KM3-230213A that could have been produced by an ultra high energy cosmic neutrino. The origin of this event is still unclear. In particular, the cosmogenic neutrino scenario is not favoured due to the non-observation of a similar event by the IceCube detector, and most galactic scenarios are disfavoured as well. We show that the blazar PKS~0605-085 is a viable source of the KM3-230213A event. In particular, even though this blazar is located at 2.4$^{\circ}$ from the KM3-230213A event, the association between the blazar and the event is not unlikely due to a sizable direction systematic uncertainty of $\approx 1.5^{\circ}$ reported by the KM3Net Collaboration. Furthermore, we show that the observation of a $\approx$72~PeV neutrino from PKS~0605-085 is entirely possible given that a $\approx$7.5~PeV neutrino could have been observed from another blazar TXS~0506+056. Finally, we consider $\gamma$-ray constraints on the number of observable neutrino events and show that for the case of the external photon field production mechanism these constraints could be relaxed due to the often-neglected effect of the isotropisation of the hadronically-produced electrons in the magnetic field of the blob. We encourage further multi-wavelength observations of the blazar PKS~0605-085.
\end{abstract}
\begin{keywords}
KM3-230213A neutrino event --- radiation mechanisms: non-thermal --- galaxies: active --- galaxies: nuclei --- quasars: individual: PKS 0605-085
\end{keywords}

\section{Introduction}

The IceCube Collaboration had reported the discovery of the astrophysical neutrino flux more than ten years ago \citep{Aartsen2013a,Aartsen2013b,Aartsen2014}. Since then, the IceCube Collaboration had, as well, reported the evidence for several neutrino sources of various types, namely, the blazar TXS~0506+056 \citep{Aartsen2018a,Aartsen2018b}, the active galactic nucleus NGC~1068 \citep{Abbasi2022} and the Galactic plane \citep{Abbasi2023}. However, a significant part of the IceCube astrophysical neutrino flux is still unassociated with any known cosmic source. 

Very recently, the KM3Net Collaboration has reported on the observation of a remarkable event KM3-230213A that could have been produced by an ultra high energy astrophysical neutrino \citep{Aiello2025} (hereafter, A25). More specifically, a through-going muon with an estimated energy of 120$^{+110}_{-60}$~PeV was detected with the ARCA array, and the energy of the parent neutrino was found to be between 72~PeV and 2.6~EeV (90 \% C.L.). The origin of KM3-230213A is still unclear; however, the authors of A25 consider four possible origins of the event, namely, the Galactic, local Universe, transient and extragalactic ones. The Galactic hypothesis is disfavoured: the authors of \citet{Adriani2025a} conclude that ``if the event is indeed cosmic, it is most likely of extragalactic origin''. Furthermore, the cosmogenic neutrino \citep{Beresinsky1969,Engel2001,Kalashev2002,Semikoz2004,Ave2005} hypothesis (and, indeed, any steady isotropic source of the KM3-230213A event) is not favoured either \citep{Li2025} since the IceCube detector which has a greater effective area (see Fig.~4 of \citet{Adriani2025b}) and observation time than the ARCA array did not detect any event(s) similar to KM3-230213A \citep{IceCube2025}.

Therefore, one is tempted to consider a discrete source (i.e. a point-like or a slightly extended one with an angular radius less than several $^{\circ}$) of an extragalactic nature as a likely source of the KM3-230213A event. In this {\it Letter} we inquire whether the blazar PKS~0605-085 (redshift $z = 0.870$, \citet{Shaw2012}) could have produced the KM3-230213A event and find such a contingency likely. In Section~\ref{sec:pks} we briefly describe the source PKS~0605-085 itself. In Section~\ref{sec:correlation} we discuss the possibility of an angular correlation between the blazar PKS~0605-085 and the KM3-230213A event. In Section~\ref{sec:qualitatively} we present a qualitative discussion and find that one neutrino event with the measured characteristics could indeed have been produced in PKS~0605-085. We deal with $\gamma$-ray constraints in Section~\ref{sec:gamma} and conclude in Section~\ref{sec:conclusions}.

\section{The blazar PKS~0605-085} \label{sec:pks}

The blazar PKS~0605-085, also known as the 4FGL~J0608.0-0835 Fermi-LAT \citep{Atwood2009} source, is a flat-spectrum radio quasar (FSRQ) located at (RA,DEC) = (92.0$^{\circ}$,-8.6$^{\circ}$) \citep{Abdollahi2020,Abdollahi2022}. It has the estimated lower-energy spectral energy distribution (SED = $E^{2}dN/dE$) peak frequency at $F_{pl} = 10^{13.7}$~Hz corresponding to the energy of $E_{pl} = 0.207$~eV \citep{Yang2022} (hereafter Y22). According to Y22, the luminosity of this peak is $L_{pl} = 10^{46.28}$~erg/s, the $\gamma$-ray luminosity is $L_{\gamma} = 10^{46.29}$~erg/s, and the X-ray luminosity is $L_{X} = 10^{45.17}$~erg/s. The KM3-230213A event was detected during a $\sim$year-long $\gamma$-ray flare of the blazar PKS~0605-085 \citep{KM3NeT2025b}.

\section{Angular correlation between the blazar PKS~0605-085 and the KM3-230213A event} \label{sec:correlation}

The KM3-230213A event was located at (RA,DEC) = (94.3$^{\circ}$,-7.8$^{\circ}$) (A25). The blazar PKS~0605-085 is located aproximately at 2.4$^{\circ}$ from the KM3-230213A event \citep{KM3NeT2025b}. Given that the estimated muon direction statistical uncertainty for the KM3-230213A event is only 0.12$^{\circ}$ (A25), one could argue that the angular distance between the blazar PKS~0605-085 and the KM3-230213A event disqualifies the blazar PKS~0605-085 from the list of possible sources of the event. However, A25 contains another crucial piece of information, namely, the total muon direction uncertainty is 1.5$^{\circ}$ and it is ``dominated by the present systematic uncertainty on the absolute orientation of the detector''. Therefore, the angular association between the blazar PKS~0605-085 and the KM3-230213A event is not unlikely.

\section{Qualitative discussion} \label{sec:qualitatively}

Let us discuss briefly what the current knowledge of neutrino production phenomenology in blazars could tell us about the likelihood of detecting a KM3-230213A-like event from the blazar PKS~0605-085. Let us use the blazar TXS~0506+056, a relatively well-established neutrino source \citep{Aartsen2018a,Aartsen2018b}, as our ``Rosetta Stone''.

\subsection{TXS~0506+056: neutrino emission and relevant parameters}

The blazar TXS~0506+056 revealed two distinct episodes of neutrino emission, in 2017 and in 2014--2015. In what follows, we will be concerned with the first episode (the 2017 one), leaving the second one for future studies. The relevant parameters of the blazar TXS~0506+056 are as follows: $z = 0.3365$ \citep{Paiano2018}, and according to Y22: $F_{pl} = 10^{14.3}$~Hz, $L_{pl} = 10^{45.81}$~erg/s, $L_{\gamma} = 10^{45.72}$~erg/s, and $L_{X} = 10^{44.19}$~erg/s. Just one neutrino event was detected from the 2017 neutrino emission episode of the blazar TXS~0506+056 via a muon track. For the assumed power-law neutrino spectrum index of 2.0 the neutrino energy is from 200~TeV to 7.5~PeV (90 \% C.L.) \citep{Aartsen2018a}.  

\subsection{Neutrino production in PKS~0605-085 vs. TXS~0506+056}

\subsubsection{Luminosity distances and the neutrino spectra}

Two principal factors work ``in favour'' of TXS~0506+056 (increasing the expected number of the neutrino events w.r.t. PKS~0605-085), namely, the smaller luminosity distance of TXS~0506+056 $d_{L} = 1.835$~Gpc vs. $d_{L} = 5.732$~Gpc for PKS~0605-085 (the $d_{L}$ ratio PKS~0605-085/TXS~0506+056 being $\approx 3$), and (at least) about one order of magnitude smaller neutrino energy (i.e. 7.5~PeV for TXS~0506+056 vs. at least 72~PeV for PKS~0605-085, the host galaxy rest frame neutrino energies are $E_{\nu-obs}(1+z) =$10.0~PeV and 135~PeV, respectively).

Let us assume the power-law index of the neutrino spectrum $dN_{\nu}/dE_{\nu} \propto E_{\nu}^{-\gamma}$ $\gamma = 2$ in both cases (we note, however, that the neutrino spectra may have low-energy and high-energy cutoffs in both cases, and the energies of these may be different). There are some theoretical indications that the neutrino spectrum from blazars around 10~PeV may have $\gamma < 2$ (i.e. it is harder than assumed here); however, we will conservatively assume $\gamma = 2$ since for $\gamma < 2$ the number of the higher-energy ($E_{\nu} > 10$~PeV) neutrino events would increase. Therefore, the mean expected number of the detected neutrino events per decade of the energy scales approximately as $N_{\nu-avg} \propto d_{L}^{-2}E_{\nu}^{-1}$, since $N_{\nu-avg} \propto (dN_{\nu}/dE_{\nu}) \Delta E_{\nu}$, where $\Delta E_{\nu}$ is the width of the energy interval in question; this gives the factor of $\approx 100$ ``in favour'' of TXS~0506+056.

\subsubsection{Effective areas of the ARCA and IceCube arrays}

The next factor worth considering is the difference of the effective areas between the ARCA and IceCube arrays. At any given energy, the ARCA effective area is smaller than the IceCube one; however, the ARCA effective area at $\approx 100$~PeV is approximately the same as the IceCube one at $\approx 10$~PeV (again, see Fig.~4 of \citet{Adriani2025b}).

\subsubsection{External photon field intensity}

For the most of the present paper, we will be concerned with the external photon field neutrino production mechanism in blazars \citep{Ghisellini2005,Ansoldi2018}, leaving the SSC mechanism (e.g. \citet{Gao2018,Cerruti2018}) for future studies. The precise mechanism by which the external photon field is produced is unknown; the spine-sheath jet structure \citep{Ansoldi2018} is among the most likely possibilities. The X-ray luminosity $L_{X}$ is a good measure of the sheath component radiation intensity (see Fig. 2 (top) of \citet{Ansoldi2018}). Thus, we conclude that the external photon field could be $10^{45.17}/10^{44.19} \approx 10$ times more luminous in PKS~0605-085 compared to TXS~0506+056. The sheath component could acquire sub-relativistic or even mildly relativistic velocities (the Lorentz factor of two or even several); this, however, does not change our conclusions qualitatively. Thus, we already have the boost factor $\times$10 in favour of PKS~0605-085 from the external photon field intensity factor alone.

\subsubsection{Transformation of external photon field to the blob rest frame, the beaming patterns and the threshold effects}

The observed neutrino intensity is influenced, among others, by the three following effects: 1) in order to produce the neutrinos, the photopion threshold must be exceeded (see e.g. eq. (15) of \citet{Kelner2008}); 2) when transformed to the blob rest frame, the energy density of the external photon field rises dramatically (see e.g. \citet{Dermer2002}); 3) finally, for the case of the external photon field the beaming pattern is different from the case of the SSC mechanism (for the external Compton $\gamma$-ray production mechanism, e.g. for a leptonic scenario, this was demonstrated in \citet{Dermer1995}). The exact impact of these three factors when combined together under the most realistic conditions is still debated \citep{Boettcher2023}; therefore, let us consider the very essence of the beaming effect, even if the resulting model will be incomplete. Here we strive to achieve simplicity and clarity rather than to develop a full theory of how the aforementioned factors tend to influence the observed neutrino intensity. Let us assume $\Gamma = D$ for simplicity, where $\Gamma$ is the blob bulk motion Lorentz factor, $D$ is the Doppler factor.

The solid angle for the jet in the observer rest frame is smaller than in the blob rest frame by a factor of $\Gamma^{2}$. Assuming that $\Gamma$ is $10^{0.5}$ times greater for PKS~0605-085 (anyway, the source of the KM3-230213A event must be exceptional due to the non-observation of a similar event with IceCube!) we get the observable neutrino intensity boost factor $\times$10. Together with the factor $\times$10 coming from $L_{X}$, the factor coming from the ratios of $\Gamma$ already gives the combined factor of $\times$100, negating the two above-described factors ``in favour'' of TXS~0506+056 (coming from the distance and the neutrino energy) completely. We conclude that if one neutrino event with the energy of 7.5~PeV could have been observed from TXS~0506+056 (this is entirely possible), then another neutrino event with the energy of 72~PeV could very well be detected from the blazar PKS~0605-085 at least if the extreme values of the relevant parameters are assumed.

Finally, what if the flare of TXS~0506+056 was more ``extreme'' (greater observed intensity ratio w.r.t. the quescent state) compared to that of PKS~0605-085? The observations show, however, that the ratio of the flaring to the quescent intensity averaged over the flare of TXS~0506+056 is barely one order of magnitude in the high energy ($E >100$~MeV) range \citep{Aartsen2018a}. Therefore, the size of this factor is less than one order of magnitude considering that PKS~0605-085 was still in the $\gamma$-ray flaring state during the detection of the KM3-230213A event. Even if the expected number of the KM3-230213A-like neutrino events from PKS~0605-085 drops from unity by the factor of 10, it is still 0.1, and PKS~0605-085 still keeps its place among the best possible sources of the event.

\section{\texorpdfstring{$\gamma$}{g}-ray constraints on the neutrino intensity} \label{sec:gamma}

An important factor constraining the observable neutrino intensity from blazars is the intensity of $\gamma$-rays from the same source. Such constraints are typically derived for the case of the SSC mechanism (see Fig.~3 of \citet{Gao2018}). In the latter case the angular distributions of the produced neutrinos and $\gamma$-rays are similar, and the transformation of the SED from the blob rest frame to the host galaxy rest frame contains the principal factor $\propto D^{4}$ (e.g. \citet{Boettcher2019,JimenezFernandez2020}). For the external photon field scenario, however, the situation is different: the produced neutrinos escape immediately, while the hadronically produced electrons and positrons keep trapped in the magnetic fields of the source producing secondary $\gamma$-rays. $\gamma$-rays may produce secondary electron-positron pairs or escape, depending on the $\gamma$-ray energy and the distance from the blob's edge. The escaped component of $\gamma$-rays is getting absorbed in the blob's environment or escape into the intergalactic medium where they may initiate intergalactic electromagnetic cascades. A part of the protons accelerated in the blob may escape the blob as well due to their relatively large interaction timescale values (compared to electrons and positrons), initiating intergalactic electromagnetic cascades as well. A detailed treatment of the intergalactic electromagnetic cascades both for the primary $\gamma$-rays and protons was presented in \citet{Dzhatdoev2017} using the ELMAG code \citep{Kachelriess2012} (see also \citet{Khalikov2021} where the deflection of the protons on intergalactic filaments was accoutted for and \citet{Blytt2020} for a newer version of the ELMAG code).

Let us assume the value of the Doppler factor $D = 100$ again. The hadronically-produced electrons and positrons (hereafter simply ``electrons'') inside the blob have the typical energy of $E_{\nu}/D \approx 1$~PeV in the rest frame of the blob. For such energies, synchrotron losses typically dominate due to the KN effect in inverse Compton scattering. The typical synchrotron photon energy [eV] (in the blob rest frame) is (e.g. \citet{Khangulyan2019}):
\begin{equation}
E_{s} = 6 \times 10^{4} \left( \frac{B}{1 \: G} \right) \left( \frac{E_{e}}{1 \: TeV} \right)^{2}, \label{eq1}
\end{equation}
where $B$ is the magnetic field, and $E_{e}$ is the electron energy, both in the blob rest frame. Assuming $B = 1$~G, we obtain $E_{s} = 6 \times 10^{10}$~eV = 60~GeV. Assuming a relatively high neutrino production efficiency (characterised by the effective proton-$\gamma$-ray optical depth $\tau_{p\gamma} = 0.1$) and an external photon field with the power-law index of $\gamma = 2$, we estimate the internal $\gamma$-ray horizon energy to be 1--10~GeV. In this case, most of the first-generation synchrotron $\gamma$-rays produced by the electrons of hadronic nature may be absorbed, and internal electromagnetic cascade develops. The energy of the second-generation electrons will be much lower than for the first one. The electron isotropisation time will be much lower than the synchrotron loss time for these second-generation electrons. If this is the case, the $\gamma$-ray constraints on the neutrino intensity will be significantly relaxed, as the neutrinos will keep their very sharp angular distribution characteristic for the case of the external photon field production mechanism \citep{Ansoldi2018}, while the associated $\gamma$-rays from the internal cascade will have a much wider angular distribution typical for the SSC scenario. If the observer is looking near the axis of the jet, the neutrino intensity is boosted w.r.t. the internal cascade $\gamma$-ray intensity. Estimates show that this boost factor may be $\times \Gamma^{2} \approx D^{2}$ or even greater; for our case, a factor of $\approx 10$ is achievable. Therefore, the $\gamma$-ray constraints could be relaxed, and we again arrive to the conclusion that one $\approx$100 PeV neutrino event from the blazar PKS~0605-085 could be detected in the ARCA array if we believe that IceCube could have indeed detected a 7.5~PeV neutrino from the blazar TXS~0506+056.



\section{Conclusions} \label{sec:conclusions}

Notwithstanding a significant angular separation of the blazar PKS 0605-085 from the KM3-230213A event (2.4$^{\circ}$), the blazar PKS 0605-085 is still a viable source of the KM3-230213A event due to the sizable systematic uncertainty of the measured neutrino direction. Model-wise, it is difficult to exclude the hypothesis that the blazar PKS 0605-085 have produced the KM3-230213A event; we personally believe that this hypothesis is, at least, not unlikely. Therefore, further multi-wavelength observations of the blazar PKS~0605-085 and futher theoretical deliberations on the subject are greatly encouraged.

\section*{Acknowledgements}

We are grateful to E.I. Podlesnyi, Prof. G.I. Rubtsov, Prof. S.V. Troitsky, and other colleagues for useful discussions. This work is supported in the framework of the State project ``Science'' by the Ministry of Science and Higher Education of the Russian Federation under the contract 075-15-2024-541.



\bibliographystyle{mnras}
\bibliography{PKS0605-085} 
\bsp	
\label{lastpage}
\end{document}